\documentclass[aps,prd,twocolumn,
eqsecnum,showpacs,nofootinbib]{revtex4}

\usepackage{amsmath,amsfonts,amssymb,color,graphicx,latexsym,theorem,mathrsfs
}
\usepackage[dvipdfmx]{hyperref}

\newcommand{\ma}[1]{\mbox{$\mathcal{#1}$}}
\newcommand{\mas}[1]{\mbox{$\mathscr{#1}$}}

\newcommand{\D}{{\rm d}}
\newcommand{\I}{{\rm i}}
\newcommand{\ti}{\tilde}
\newcommand{\we}{\wedge}

\begin{document}

\title{
Euclidean supersymmetric solutions with the self-dual Weyl tensor
}

\author{Masato Nozawa}
\email{
masato.nozawa@yukawa.kyoto-u.ac.jp
}


\address{ 
Yukawa Institute for Theoretical Physics and Department of Physics, Kyoto University, Kyoto 606-8502, Japan
}

\date{\today}

\begin{abstract} 
We explore the Euclidean supersymmetric solutions admitting the self-dual gauge field in the framework of ${\cal N}=2$ minimal gauged supergravity in four dimensions. According to the classification scheme utilizing the spinorial geometry or the bilinears of Killing spinors, the general solution preserves one quarter of supersymmetry and is described by the Przanowski-Tod class with the self-dual Weyl tensor. We demonstrate that there exists an additional Killing spinor, provided the Przanowski-Tod metric admits a Killing vector that commutes with the principal one. The proof proceeds by recasting the metric into another Przanowski-Tod form. This formalism enables us to show that the self-dual Reissner-Nordstr\"om-Taub-NUT-AdS metric possesses a second Killing spinor, which has been missed  over many years. We also address the supersymmetry when the Przanowski-Tod space is conformal to each of the self-dual ambi-toric K\"ahler metrics. It turns out that three classes of solutions are all reduced to the self-dual Carter family, by virtue of the nondegenerate Killing-Yano tensor. 
\end{abstract}

\maketitle


\section{Introduction}
The localization technique~\cite{Pestun:2007rz,Pestun:2016zxk}  in supersymmetric gauge theories defined on a curved Riemannian background provides a powerful tool to implement exact non-perturbative calculations, such as the expectation value of a Wilson loop and the partition functions. The localization principle therefore is a valuable quantitative instrument to explore the strongly coupled regime, as worked out in $\mathcal N=2 $ supersymmetric gauge theories in three dimensions~\cite{Hama:2011ea,Imamura:2011wg}. In line with this progress, there has been an increasing interest in the study of gravitational solutions with Euclidean signature in dual gravity theories. Some gravitational dual solutions were studied in detail and exploited to reproduce elegantly the field theory results. See e.g., \cite{Martelli:2011fu,Martelli:2012sz,Martelli:2013aqa,Farquet:2014kma} and references cited therein.  

On account of the $G$-structure restriction coming from the Killing spinor, 
the corresponding supersymmetric solutions to the four-dimensional Euclidean gauged supergravity can be classified in a systematic manner~\cite{Dunajski:2010zp,Dunajski:2010uv,Dunajski:2013qc} (see also \cite{Klemm:2015mga}).  For the minimal gauged supergravity in four dimensions, it turned out that the system reduces to the nonlinear equations for two variables on the three-dimensional curved space, and the general solution preserves one quarter of supersymmetry, provided the Killing spinor takes the same form as in the Lorentzian counterpart. Among them, an interesting subclass arises when the gauge field is self-dual, for which the general metric is described by the Einstein space with a self-dual Weyl curvature found by Przanowski~\cite{Przanowski:1991ru}  and Tod~\cite{Tod:2006wj}. The Einstein metrics with a self-dual Weyl tensor have been studied in a variety of contexts including twistor spaces and gravitational instantons.

The Przanowski-Tod metric is conformal to the scalar-flat K\"ahler space found by LeBrun~\cite{LeBrun}. 
The solution allows a Killing vector which preserves the integrable complex structure and is specified by a single function obeying the continuous Toda equation. Despite the fact that the governing system is remarkably simplified compared to Einstein's equations, finding the exact solutions to the nonlinear Toda equation is still a formidable task, which is a main obstacle to the comprehensive survey of supersymmetric solutions. Furthermore, we do not know {\it a priori} what kind of solution to the continuous Toda equation provides the gravitational solution of mathematical and physical interest.  An astute strategy to evade these difficulties is to  find the coordinate transformation which converts the solutions of interest into the Przanowski-Tod form.  In the recent joint work with Houri~\cite{Nozawa:2015qea},  the present author worked out the necessary and sufficient conditions for the Euclidean supersymmetry of the most general Petrov-D solution found by Pleba\'nski and Demia\'nski~\cite{Plebanski:1976gy}.  Ref.~\cite{Nozawa:2015qea} proved that the self-dual Pleba\'nski-Demia\'nski solution admits two independent Killing spinors, whereas the non-self dual solution admits a single Killing spinor. The method embraced there was to cast the Pleba\'nski-Demia\'nski  metric into two different Przanowski-Tod forms, by making full use of the Killing-Yano tensor found in \cite{Houri:2014hma} 
 and the ambi-K\"ahler property~\cite{Apostolov:2013oza}. This was the first work that demonstrated explicitly that the self-dual solution admits the enhancement of super and hidden symmetries, compared to the non-self-dual counterparts. 

Nevertheless, ref.~\cite{Nozawa:2015qea} left several issues open to debate. 
Since the nondegeneracy of the Killing-Yano tensor has been postulated in~\cite{Nozawa:2015qea}, it has remained unclear if the self-dual Reissner-Nordstr\"om-Taub-NUT-AdS solution, in which the Killing-Yano tensor is degenerate, allows a second Killing spinor. 
In addition, the Killing-Yano tensor for the self-dual Pleba\'nski-Demia\'nski solution was constructed 
based on the Petrov-D property. This makes it obscure whether the presence of the second Killing spinor for the self-dual metric is generic or specific to the algebraically special solutions. To address these unsettled issues is one of the main purposes of the present article. 

For the Euclidean Reissner-Nordstr\"om-Taub-NUT-AdS solution, 
one may be tempted to hope that the Killing spinor equation might be integrated directly, 
since the solution enjoys a number of symmetries. In the self-dual case, however, one immediately encounters serious difficulties in this approach. As pointed out in \cite{Nozawa:2015qea}, the first integrability condition of the Killing spinor necessarily acquires two zero eigenvalues. This fact forbids us to extract a useful projection operator for integration. Furthermore,  the self-dual gauge field appearing in the Killing spinor equation disagrees in general  with the self-dual limit of the gauge field of non-self-dual solutions~\cite{Farquet:2014kma,Nozawa:2015qea}. It follows that one cannot work out the (non)existence of Killing spinors, as far as one is adhering to the self-dual limit of the non-self-dual gravitational solutions. This problem has plagued previous studies and some confusions have prevailed as to the correct fraction of preserved supersymmetry of the self-dual solutions. 

In this paper, we undertake the elaborated analysis on the supersymmetry of the Przanowski-Tod metric in the framework of ${\cal N}=2$ minimal gauged supergravity. A key ingredient here is the additional Killing vector which is linearly independent of the principal Killing vector.  When the Przanowski-Tod space is independent of the coordinate $y$ (see equation~(\ref{LeBrun}) below), it follows that the metric can be brought into the Calderbank-Pedersen form~\cite{Calderbank:2001uz}, as first demonstrated in \cite{Ward:1990qt}. We show that the Przanowski-Tod metric allows a one-parameter family of different description. Since any Przanowski-Tod metric admits a Killing spinor of one quarter of supersymmetry, this means that two independent Killing spinors exist. 
This is to be contrasted with the conventional supersymmetry enhancement, because the gauge field in the Killing spinor equation is distinct in the present setting from the original one and is also characterized by a single parameter.

The remainder of the  paper constitutes as follows. The next section presents a simple formulation how the 
self-dual Przanowski-Tod metric admits the second independent Killing spinor. In section~\ref{sec:RNAdS}, we use this framework to establish that the self-dual Reissner-Nordstr\"om-Taub-NUT-AdS solution admits one more Killing spinor, in contrast to the claim in the literature that this metric preserves only one quarter or none of supersymmetry. In section~\ref{sec:ambiKahler}, supersymmetry of three classes of the conformal ambi-K\"ahler metrics is explored. In the self-dual case, all three classes of the conformal ambi-toric K\"ahler metrics are degenerate into the self-dual Carter family~\cite{Carter:1968ks}, on account of the existence of the  nondegenerate Killing-Yano tensor in each class.  We draw our conclusions in the final section \ref{remark} with several future prospects.

\section{Supersymmetry of the self-dual solution}
\label{sec:sd}

Let us consider the Euclidean Einstein-Maxwell  system with a negative cosmological constant, whose action is given by
\begin{align}
\label{action}
S=-\frac 12 \int (R+6\ell^{-2})\star 1-2F \we \star F \,. 
\end{align}
Here $\ell $ denotes the (Euclidean) AdS radius with its reciprocal being the gauge coupling constant. 
$F$ is the ${\rm U}(1)$ field strength corresponding to the graviphoton and can be locally expressed by $F=\D A$. 
The gravitational solution is said to preserve supersymmetry, provided it admits a nontrivial spinor 
$\epsilon$ obeying the 1st-order differential equations
\begin{align}
\label{KS}
\hat \nabla_\mu \epsilon\equiv \left(\nabla_\mu+
\frac{\I}{4}F_{\nu\rho}\gamma^{\nu\rho}\gamma_\mu+\frac{1}{2\ell}\gamma_\mu
-\frac{\I}{\ell }A_\mu \right)\epsilon=0\,.
\end{align}
$\epsilon$ is the ${\rm Spin}(4)$ Dirac spinor and $\gamma_\mu $ defines the 
Clifford algebra $\{\gamma_\mu, \gamma_\nu\}=2g_{\mu\nu}$. 
The general supersymmetric solutions have been classified systematically in
\cite{Dunajski:2010uv,Nozawa:2015qea} and  preserve at least one quarter of supersymmetry
when $A$ and $\ell$ are both real.

Let us focus hereafter on the case in which 
the Maxwell field is self-dual
\begin{align}
\label{Fsd}
F=\star F \,.
\end{align}
 In this case, 
the metic is Einstein $R_{\mu\nu}=-3\ell^{-2} g_{\mu\nu}$ because the stress-energy tensor
vanishes identically, i.e., $F$ denotes an instanton. 
The general analysis in \cite{Dunajski:2010zp,Nozawa:2015qea} reveals that 
the local metric of supersymmetric solutions with a self-dual gauge field 
can be written in the Przanowski-Tod form~\cite{Przanowski:1991ru,Tod:2006wj}
\begin{align}
\label{PT}
\D s^2=\frac{\ell^2}{z^2}\D \hat s^2_{\rm LeBrun}\,,
\end{align}
where
\begin{align}
\label{LeBrun}
\D \hat s^2_{\rm LeBrun}=H^{-1}(\D t+\omega)^2+H[\D z^2+e^u (\D x^2+\D y^2)]\,, 
\end{align}
and\footnote{
Since the length scale $\ell$ has been factored out in (\ref{PT}), 
the expressions (\ref{H}), (\ref{omega}) are obtained by simply setting $\Lambda=-1/2$ in \cite{Dunajski:2010zp,Tod:2006wj}. 
}
\begin{align}
\label{H}
H&=1-\frac 12 zu_z \,,\\
\label{omega}
\D \omega &=H_x\D y \we \D z-H_y\D x\we \D z+(e^u H)_z\D x\we \D y\,. 
\end{align}
Here and throughout the paper, we use the notation 
$u_x=\partial u/\partial x$ etc to denote the partial differentiation. 
The solution is independent of $t$ and entirely controlled by a single function 
$u=u(x,y,z)$ satisfying the continuous Toda equation
\begin{align}
\label{Toda}
u_{xx}+u_{yy}+(e^u)_{zz}=0\,.
\end{align}
The integrability $\D^2 \omega=0$ in  (\ref{omega}) is satisfied by the Toda equation (\ref{Toda}). 
The Przanowski-Tod metric is complex and hermitian, and its Weyl tensor is self-dual
\begin{align}
\label{sdWeyl}
C_{\mu\nu\rho\sigma}=\frac 12 \epsilon_{\mu\nu\tau\lambda}C^{\tau\lambda}{}_{\rho\sigma}\,.  
\end{align}
The self-duality of the Weyl tensor comes from the constraints of supersymmetry. 
The four-dimensional Einstein manifold with a self-dual Weyl tensor possesses a quaternionic K\"ahler structure~\cite{Besse}.

Let us digress a bit here and devote ourselves to the conformally transformed metric $\D \hat s_{\rm LeBrun}^2=(z/\ell)^2 \D s^2$. 
This metric describes the LeBrun space~\cite{LeBrun}, 
which is the scalar-flat K\"ahler  manifold with an anti-self-dual
K\"ahler form $\hat \Omega=(\D t+\omega)\we \D z -H e^u \D x\we \D y$. 
Here the hat notation is intended to highlight the quantities defined on the LeBrun space and therefore
their indices are raised and lowered by the LeBrun metric and its inverse. 
The coordinate $z$ plays the role of the conformal factor, as well as the 
the moment map of the LeBrun space. The Ricci form 
$\hat{\ma R}_{\mu\nu}=\frac 12 \hat \Omega_{\rho\sigma}\hat R^{\rho\sigma}{}_{\mu\nu}$
of the LeBrun space is locally expressed by $\hat{\ma R}=\D \hat{\cal P}$, where 
\begin{align}
\label{RicciF}
\hat{\cal P}= \frac{u_z}{2H}(\D t+\omega)+\frac 12 (-u_y \D x+u_x\D y) \,. 
\end{align}
(\ref{omega}) assures that the  K\"ahler form $\hat \Omega$ is covariantly constant
and (\ref{Toda}) imposes the self-duality of the Weyl curvature\footnote{
In contrast, (\ref{H}) is the condition which assures the Przanowski-Tod metric to be 
the Einstein space.  Note also that the self-duality conditions (\ref{Fsd}), (\ref{sdWeyl})  are conformally invariant and therefore valid for both metrics. } 
 which amounts to the 
vanishing of the scalar curvature $\hat R=0$. For the LeBrun space,  $H$ and $u$ are independent functions, thereby
$H$ need not satisfy (\ref{H}). 
The governing equation for $H$ is obtained by the integrability $\D^2\omega=0$, yielding
\begin{align}
\label{HLeBrun}
H_{xx}+H_{yy}+(e^uH)_{zz}=0 \,. 
\end{align}

Let us now turn our attention back to the Killing spinor of the Przanowski-Tod space. 
The gauge field appearing in the Killing spinor is given by~\cite{Dunajski:2010uv,Farquet:2014kma,Nozawa:2015qea}
\begin{align}
\label{Finst}
A=-\frac{\ell}{2}\hat{\cal P}\,, 
\end{align}
where $\hat{\cal P}$ is the Ricci form gauge potential (\ref{RicciF}) of the LeBrun space. 
Namely, $A$ denotes a connection on the ${\rm spin}^c$  bundle of the K\"ahler manifold~\cite{Farquet:2014kma}. 
Taking the orthonormal frame
\begin{align}
\label{}
&e^1=\frac{\ell}{z}H^{-1/2}(\D t+\omega)\,, \qquad e^4=\frac{\ell}{z}H^{1/2}\D z\,, \notag\\ 
&e^2=\frac{\ell}{z}H^{1/2}e^{u/2}\D x \,, \qquad e^3=\frac{\ell}{z}H^{1/2}e^{u/2}\D y\,,
\end{align} 
and inserting (\ref{Finst}) into  (\ref{KS}), 
the Killing spinor equation is integrated to give~\cite{Farquet:2014kma,Nozawa:2015qea}
\begin{align}
\label{KSPT}
\epsilon=\frac 14\sqrt{\frac{\ell}z}(1-i H^{-1/2}\gamma^1 )
(1-i \gamma^{23})(1+\gamma_5)\epsilon _0 \,,
\end{align}
where $\gamma_5=\gamma_{1234}$  and $\epsilon_0$ is a constant Dirac spinor.  
Because of the two independent projection operators in (\ref{KSPT}), the Przanowski-Tod
space preserves at least one quarter of supersymmetry. 
The bilinear vector 
$V^\mu =i\epsilon^\dagger \gamma_5\gamma^\mu \epsilon=(\partial/\partial t)^\mu $ defines a principal Killing vector which keeps invariant the metric 
$\mas L_V g_{\mu\nu}=0$ and the complex structure $\mas L_VJ=0$.  

As shown in~\cite{Nozawa:2015qea}, the matrices 
$[\hat \nabla_\mu, \hat \nabla_\nu]$ are proportional to $(1-\gamma_5)$ for the 
self-dual solutions of the Killing spinor (\ref{KS}). This means that ${\rm det}[\hat \nabla_\mu, \hat \nabla_\nu]$ necessarily admit at least two zero eigenvalues. This fact is indicative of the feature that half of the supersymmetry is preserved. However, the integrability is only a necessary condition~\cite{vanNieuwenhuizen:1983wu} , so that the mere requirement of the integrability condition is a lack of mathematical rigor. In order to verify the enhanced supersymmetry robustly,  one has to construct explicit Killing spinors. 

Suppose that the Przanowski-Tod metric admits another linearly independent Killing spinor $\ti \epsilon$.  
Then, its bilinear vector field $\ti V^\mu =i\ti \epsilon^\dagger \gamma_5\gamma^\mu\ti \epsilon$
is also a Killing vector which is linearly independent of 
$V^\mu =i\epsilon^\dagger \gamma_5\gamma^\mu \epsilon$. 
It turns out that there exist two different ways of writing the metric into the Przanowski-Tod forms. Observe that 
these bilinear Killing vectors are not necessarily commutative with each other, as one can infer from the maximally supersymmetric ${\rm AdS}_2\times S^2$ solution in Lorentzian ungauged supergravity.  

In order to capture the essence of the idea, we shall concentrate here on the simple case in which the metric (\ref{PT}) admits another Killing vector of the form $ \partial/ \partial y$. In this case, one can set $\omega=\omega_0 (x,z)\D y$
without losing any generality, and (\ref{omega}) now simplifies  to 
$(\omega_0)_z=-H_x$ and $(\omega_0)_x=(e^uH)_z$. 
It turns out that the conformal K\"ahler structure must be toric. As shown in \cite{Ward:1990qt},  this class of solutions falls into the Calderbank-Pedersen family~\cite{Calderbank:2001uz}, which stands for the most general Einstein metric with the self-dual Weyl tensor possessing two linearly independent commuting Killing vector fields. We shall show shortly that this case indeed admits the second independent Killing spinor and includes a number of mathematically interesting geometries.

If the second Killing spinor $\ti \epsilon $ exists in this class of metrics, its bilinear Killing vector
$\ti V^\mu=i\ti\epsilon \gamma_5\gamma^\mu\ti\epsilon$
must be built out of $\partial/ \partial t $ and $\partial/ \partial y$ as 
\begin{align}
\label{Vtilde}
\ti V=c_1 \frac{\partial}{\partial t} +c_2 \frac{\partial}{\partial y}  \,,
\end{align}
where $c_1$ and $c_2$ are constants. 
Here, we wish to find the coordinate transformation $(t, x,y,z)\mapsto (\ti t, \ti x, \ti y , \ti z)$ 
which brings the metric (\ref{PT}) into another Przanowski-Tod form
\begin{align}
\label{PT2}
\D s^2=\frac{\ell^2}{\ti z^2}\left[
\ti H^{-1}(\D\ti  t+\ti \omega)^2+\ti H \{\D \ti z^2+e^{\ti u} (\D \ti x^2+\D \ti y^2)\}
\right]\,,
\end{align}
with $\ti V=\partial/\partial \ti t$. Here, 
$\ti H$, $\ti \omega$ and $\ti u$ should satisfy equations of the `tilded' Toda system
(\ref{H}), (\ref{omega}), (\ref{Toda}). 

A key observation to obtain the suitable transformation is the twistor tensor which is present in Einstein spaces with the self-dual Weyl curvature~\cite{Tod:2006wj} 
\begin{align}
\label{twistor}
k =\frac 12 (\D \ti V^\flat -\star \D \ti V^\flat ) \,, 
\end{align}
where $\ti V^\flat$ is a one-form dual to the Killing vector $\ti V$ (\ref{Vtilde}). The two-form $k$
is anti-self-dual and satisfies the conformal Killing-Yano equation~\cite{Tod:2006wj}. 
The desired conformal factor $\ti z$ is then found to be 
\begin{align}
\label{tildez}
\ti z =\frac{2}{\sqrt{k_{\mu\nu}k^{\mu\nu}}} =\frac{2z}{h(x,z)}\,,
\end{align}
where $h(x,z)\equiv \sqrt{4c_2^2 e^u +(2c_1+2c_2 \omega_0-c_2 z u_x)^2}$. 
After some manipulations, we find that the rest of the coordinate transformations is given by
\begin{align}
\ti t=&\frac{c_1t+c_2 y}{c_1^2+c_2^2} \,, \qquad 
\ti y=-c_2 t+c_1 y \,, \notag \\
\label{dxt}
\D \ti x=&\frac{2c_2^2 e^u H +(c_1+c_2\omega_0)(2c_1+2c_2\omega_0-c_2zu_x)}{h(x,z)}\D x
\notag \\&+\frac{c_2z[(c_1+c_2\omega_0)u_z +c_2 H u_x]}{h(x,z)}\D z \,, 
\end{align}
where the integrability $\D ^2 \ti x=0$ in (\ref{dxt})  is assured by the Toda system (\ref{H}), (\ref{omega}), (\ref{Toda}), which therefore guarantees the local existence of the coordinate $\ti x$.   
With the above new coordinates, one can bring the Przanowski-Tod metric (\ref{PT}) into 
another canonical form (\ref{PT2}) with 
\begin{align}
\ti H &=\frac{Hh(x,z)^2}{4[(c_1+c_2\omega_0)^2+c_2^2 e^u H^2]}\,,\notag \\
\ti \omega &= \frac{(c_1^2-c_2^2)\omega_0+c_1c_2(\omega_0^2-1+e^u H^2)}
{(c_1^2+c_2^2)[(c_1+c_2\omega_0)^2+c_2^2 e^u H^2]}\D \ti y\,, \label{tiHou}\\ 
e^{\ti u}&=\frac{16 e^u}{h(x,z)^4}\,. \notag 
\end{align}
Some tedious but straightforward computations show that $\ti H$, $\ti \omega$ and $\ti u$ satisfy the `tilded' Toda system (\ref{H}), (\ref{omega}), (\ref{Toda}). It turns out that the Killing spinor equation 
can be integrated in the new coordinate system ($\ti t,\ti x,\ti y,\ti z$) and the solution $\ti \epsilon$ takes the form (\ref{KSPT}) in the tilded frame with the gauge field
\begin{align}
\label{tiA}
\ti A=&-\frac{\ell [c_2 u_x H+u_z(c_1+c_2\omega_0)]}{2H h(x,z)}(\D t+\omega_0\D y) \\
&+\frac {\ell [2c_2(e^u)_z-u_x (2 c_1+2c_2\omega_0-c_2 z u_x)] }{4 h (x,z)}\D y
\,. \notag
\end{align}
Since this Killing spinor $\ti \epsilon$ is obviously independent of the first, this shows that there appears another supersymmetry generated by $\ti\epsilon$. The explicit coordinate transformations (\ref{tildez}), (\ref{dxt}) represent the  main result in this section.   This argument does not depend on the degeneracy of the Killing-Yano tensor nor the Petrov-D property, extending substantially the analysis given in~\cite{Nozawa:2015qea}. Following the terminology in \cite{Tod:2006wj}, the above coordinate transformations interchange the role of class A and B which distinguish the solutions of the Przanowski master equation~\cite{Przanowski:1991ru}.

Ref. \cite{Dunajski:2013qc} analyzed the condition under which  the Euclidean supersymmetry 
is enhanced by focusing on the integrability of the Killing spinor equation. 
They concluded that the half-supersymmetric self-dual solutions are exhausted by the one for which the 
Toda equation is separable with respect to the coordinates ($x,y$) and $z$ (see (4.33) in  \cite{Dunajski:2013qc}). 
Their analysis in \cite{Dunajski:2013qc} is consistent with the present result, since in the present case the instanton 
given in (\ref{Finst}) is inequivalent to the `tilded' one (\ref{tiA}). This possibility has not been  examined in~\cite{Dunajski:2013qc}. 
Examples considered in the following two sections manifest this feature and the Toda equation is not of the separable form.


\section{Reissner-Nordstr\"om-Taub-NUT-AdS solution}
\label{sec:RNAdS}

Exploiting the algorithm given in the previous section, we can show the 
existence of the two independent Killing spinors in the self-dual Euclidean Reissner-Nordstr\"om-Taub-NUT-AdS solution. 
The non-self-dual Euclidean Reissner-Nordstr\"om-Taub-NUT-AdS solution reads
\begin{align}
\label{RNAdS}
\D s^2=&\frac{\Delta(r)}{R(r)^2}(\D \tau -2n v \D \phi)^2\notag \\
&+R(r)^2 
\left[\frac{\D r^2}{\Delta (r)}+\frac{\D v^2}{1-v^2}+(1-v^2)\D \phi ^2 \right]\,,\\
\label{RNAdSA}
A=&\frac {q_e r-n q_m}{R(r)^2} \D \tau+\frac{q_m(r^2+n^2)-2n q_e r}{R(r)^2}v\D \phi\,,
\end{align}
with 
\begin{align}
\label{}
R(r)=&\sqrt{r^2-n^2} \,, \\
\Delta(r)=&\frac{R(r)^4}{\ell^2}+
\left(1-\frac{4n^2}{\ell^2}\right)(r^2+n^2)-2mr +q_m^2-q_e^2\,.  \notag
\end{align}
The coordinate $\cos\theta =-v$ denotes the usual azimuthal angle and 
$\phi$ is $2\pi$ periodic. 
The solution admits ${\rm U}(2)={\rm U}(1)\times {\rm SU}(2)$ symmetry and is specified by four parameters: the mass $m$, the NUT charge $n$, the electric charge $q_e$ 
and the magnetic charge $q_m$. 
Taking the orientation $\D \tau \we \D r\we \D v\we \D \phi$ to be positive, 
the gauge-field is self-dual $F=\star F$ when 
\begin{align}
\label{RNAdS_sdF}
q_e=q_m \,. 
\end{align}
The self-duality of the Weyl tensor (\ref{sdWeyl}) boils down to 
\begin{align}
\label{RNAdS_sd}
q_e=q_m\,, \qquad 
m=n-\frac{4n^3}{\ell^2} \,. 
\end{align}
When (\ref{RNAdS_sd}) is fulfilled, 
the largest root of $g_{tt}=0$ is $r=n(>0)$, at which the regularity 
of the metric requires $\tau $ to have a period $8\pi n$.

The metric (\ref{RNAdS}) follows from the appropriate scaling limit of the general 
Pleba\'nski-Demia\'nski solution~\cite{Plebanski:1976gy}. 
The necessary and sufficient conditions under which the Pleba\'nski-Demia\'nski solution (including the C-metric and the Carter family) is supersymmetric were obtained in \cite{Klemm:2013eca} for the Lorentzian signature. The supersymmetry of the Reissnerr-Nordstr\"om-Taub-NUT-AdS solution in Lorentzian signature was first addressed in \cite{AlonsoAlberca:2000cs} (see also \cite{Romans:1991nq,Caldarelli:1998hg}).
Ref. \cite{Nozawa:2015qea} studied the Euclidean supersymmetry of the general 
Pleba\'nski-Demia\'nski solution, and showed that two independent Killing spinors exist in the self-dual case  
utilizing the nondegeneracy of the Killing-Yano tensor. Our discussion here is insensitive to this property and hence applicable also to the above solution (\ref{RNAdS}) for which the Killing-Yano tensor is degenerate.

 
Following the procedure given in \cite{Klemm:2015mga,Klemm:2013eca}, we can transform the metric (\ref{RNAdS})
into the Przanowski-Tod form (\ref{PT}) by 
the coordinate transformation
\begin{align}
\label{zx_RNAdS}
z&=\frac{\ell^2}{r-n}\,, \qquad 
x=\frac 12 \log \left(\frac{1-v}{1+v}\right) \,, \notag \\
t&= \tau \,, \qquad y=\phi \,. 
\end{align}
The metric functions are then given by
\begin{align}
\label{Hou_RNAdS}
H&=\frac{r^2-n^2}{(r+n)^2+\ell^2-4n^2}\,, \qquad  \omega_0=-2nv\,, \notag \\
e^u&=\frac{\ell^2(1-v^2)}{(r-n)^2}[(r+n)^2+\ell^2-4n^2]\,. 
\end{align}
One can easily verify that all equations of the Toda system (\ref{H}), (\ref{omega}), (\ref{Toda}) 
are satisfied.  Equations (\ref{RicciF}) and (\ref{Finst}) yield the instanton gauge potential, 
which turns out to be a constant multiple of the self-dual limit of (\ref{RNAdSA}) up to gauge. 
This confirms the existence of the first Killing spinor~\cite{Dunajski:2013qc,Farquet:2014kma}. 

To get further insight, let us define $w=\ell^2(r+n)/(r-n)$ and cast the metric  
(\ref{RNAdS}) into the following form
\begin{align}
\label{}
\D s^2=&\frac{4\ell^2n^2}{(\ell^2-w)^2}\Biggl[
\frac{W(w)}{w}\left(\frac{\D \tau}{2n}-v\D \phi \right)^2+\frac{w\D w^2}{W(w)}
\notag \\&+w 
\left(\frac{\D v^2}{1-v^2}+(1-v^2)\D \phi^2 \right)
\Biggr]\,, 
\end{align}
where $W(w)=\ell^2(\ell^2-4n^2)+2(4n^2-\ell^2)w+w^2$. 
The metric in the square bracket stands for the canonical metric for the ambi-toric K\"ahler  of Calabi-type~\cite{Apostolov:2013oza}. 
It is interesting to observe that the half supersymmetric solutions of this kind also emerge when a
single instanton field is not (anti-)self-dual (see (4.27) in~\cite{Dunajski:2013qc}).

Since $\partial/\partial y$ is the Killing vector for the Euclidean Reissner-Nordstr\"om-Taub-NUT-AdS solution (\ref{RNAdS}), another Killing spinor can be obtained by following the reasoning in section~\ref{sec:sd}. The explicit forms of $\ti H, \ti \omega, \ti u$ in the new Przanowski-Tod form 
can be easily inferred by inserting (\ref{zx_RNAdS}), (\ref{Hou_RNAdS}) into (\ref{tiHou}), 
whereas the integration of $\ti x$ in (\ref{dxt}) is a bit cumbersome but nonetheless computable as\begin{align}
\label{}
\ti x=&\frac {c_1}2 \log \left(\frac{1-v}{1+v}\right)+c_2n \log \left[\frac{D_1}{(r-n)^2(1-v^2)}\right] \notag 
\\&
-\alpha c_2 \sin^{-1} 
\left[\frac{c_1D_2+c_2 \alpha^2 (r-n)v}{\ell\sqrt{(c_1^2+c_2^2\alpha ^2)D_1}}\right] \notag\\
&+\frac{c_1}{2}\log \left[
\frac{(D_3+c_2 D_2)^2+[c_2\alpha (r-n)(1+v)]^2}
{(D_3-c_2 D_2)^2+[c_2\alpha (r-n)(1-v)]^2} \right]
\notag \\
&+2 c_2n \log \left[\frac{\alpha^2 (r-n)(D_3-c_2D_2v )}
{\ell^2\sqrt{c_1^2+c_2^2\alpha^2}
D_1}\right] \,,
\end{align}
where $\alpha =\sqrt{\ell^2-4n^2}$ and 
\begin{align}
\label{}
D_1=&(r+n)^2+\alpha^2 \,, \qquad D_2=\ell^2+2n(r-n)\,, \notag \\
D_3=& \sqrt{[c_1(r-n)-c_2 D_2v]^2+c_2^2\ell^2 D_1(1-v^2)}\notag \\&+c_1(r-n)\,.
\end{align}
The instanton gauge field in the new frame reads
\begin{align}
\label{}
\ti A
=& -\frac{(r-n)[c_1D_2+c_2\alpha^2(r-n)v]}{2\ell (r+n)[D_3-c_1(r-n)]}(\D \tau-2nv\D \phi)\notag \\
&-\frac{\ell[c_1(r-n)v-c_2 D_2]}{2 [D_3-c_1(r-n)]}\D \phi ,
\end{align}
which disagrees with the self-dual limit of (\ref{RNAdSA}), as we underlined. 
Since the complexity of these expressions has been a main obstacle for integrating the Killing spinor equation in the original coordinates, the Killing spinor constructed here has been missed so far. 
 This illustrates the power of the general framework organized in section~\ref{sec:sd}. From the viewpoint of three-dimensional Chern-Simons theories, the ratio $c_2/c_1$ corresponds to the choice of the almost contact structure in three sphere~\cite{Martelli:2011fu}. 


\section{Conformal ambi-toric K\"ahler metrics}
\label{sec:ambiKahler}

In this section, we study the Przanowski-Tod metric in which the  LeBrun space is taken to the regular type ambi-toric K\"ahler.  As well as illustrating our general results in section~\ref{sec:sd}, 
this example allows us to discover the unexpected hidden symmetry represented by the Killing-Yano tensor.

An ambi-K\"ahler structure comprises a pair of K\"ahler structures 
($\hat g_\pm , \hat \Omega_\pm, \hat J_\pm$) 
for which the metrics are conformally related $\hat g_-=\Phi^2 \hat g_+$ for some function $\Phi$ with the opposite orientation
$\hat \Omega_+\we \hat \Omega_+=-\hat \Omega_-\we \hat\Omega_-$~\cite{Apostolov:2013oza}. 
When these K\"ahler metrics are both toric, the analysis in~\cite{Apostolov:2013oza} clarified that 
the ambi-toric K\"ahler geometries are either of Calabi-type or regular,  and the latter class is divided into three varieties (hyperbolic, parabolic and elliptic types). Since the self-dual Reissner-Nordstr\"om-Taub-NUT-AdS solution belongs to the conformal K\"ahler of Calabi-type as verified in the previous section, we direct attention here toward the 
regular type. In terms of the barycentric metric 
$\D s_c^2=\Phi^{-1}\D \hat  s^2_-=\Phi\D \hat  s^2_+$ of the form
\begin{align}
\label{}
\D s_c^2=\D \hat s_2^2 +\frac{\D q^2}{Q(q)}+\frac{\D p^2}{P(p)}\,, 
\end{align}
where $P(p)$, $Q(q)$ are arbitrary functions, 
these three subclasses of regular type are specified as follows. (i) {\it Hyperbolic}
\begin{align}
\D  \hat s_{2}^2 =&
\frac{Q(q)(\D \tau+p^2 \D \sigma)^2}{(q^2-p^2)^2}
+\frac{P(p)(\D \tau+q^2 \D \sigma)^2}{(q^2-p^2)^2}\,,\notag \\
\hat \Omega_\pm =& \D \left(\frac{\D \tau}{q\mp p}\pm \frac{pq}{q\mp p}\D \sigma\right) \,, \quad 
\Phi=\frac{q-p}{q+p}\,. \label{hyperbolic}
\end{align}
(ii) {\it Parabolic}
\begin{align}
\D\hat  s_2^2=&
\frac{Q(q)(\D \tau+p \D \sigma)^2}{(q-p)^2}
+\frac{P(p)(\D \tau+q \D \sigma)^2}{(q-p)^2}\,,\notag \\
\hat \Omega_+ =& -\D \left(\frac{\D \tau}{p-q}+\frac{p+q}{2(p-q)}\D \sigma\right)\,, \notag \\
\hat \Omega_- =& -\D [(p+q)\D \tau +p q \D \sigma]\,,\quad \Phi=q-p\,.
\label{parabolic}
\end{align}
(iii) {\it Elliptic} (this fixes the typo in~\cite{Apostolov:2013oza})
\begin{align}
\D \hat s^2_2=
&\frac{P(p)}{(q-p)^2(1+pq)^2}[2 q\D \tau+(q^2-1) \D \sigma]^2 \notag \\
&+
\frac{Q(q)}{(q-p)^2(1+pq)^2}[2p\D \tau+(p^2-1) \D \sigma]^2 \,,\notag \\
\hat \Omega_+=&\D \left(\frac{q+p}{q-p}\D \tau-\frac{1-pq}{q-p}\D \sigma\right)\,, \notag \\
\hat \Omega_-=&\D \left(\frac{2\D \tau}{1+pq}+\frac{p+q}{1+pq}\D \sigma\right)\,,\quad 
\Phi=\frac{q-p}{1+pq}\,.
\label{elliptic}
\end{align}
Here $\hat \Omega_\pm $ are the K\"ahler forms with 
$\hat \star \hat \Omega_\pm =\pm \hat \Omega_\pm$
in the positive orientation $\D \tau\we \D q \we \D p \we \D \sigma$. 
Each ambi-toric K\"ahler metric is endowed with two commuting Killing vectors 
$\partial/\partial\tau$ and $\partial/\partial\sigma$, and is characterized by two 
structure functions. We shall show below that these metrics are all incorporated into the $y$-independent LeBrun metric 
and its conformal class admits the self-dual Einstein metric of the Przanowski-Tod form. 


Since the general ambi-toric K\"ahler solutions (\ref{hyperbolic}), (\ref{parabolic}), (\ref{elliptic})
 are not self-dual nor Einstein (hence $P(p)$, $Q(q)$ are left undetermined), there  exist diverse ways of transformations from the non-self-dual LeBrun to each ambi-toric K\"ahler solution
$(t,x,y,z)\mapsto (\tau, q,p,\sigma)$. However, our interest here lies in the appropriate $z$ which gives rise to the Przanowski-Tod form via the conformal transformation. Hence, we present only the bottom line of $z$ which indeed realizes that the metric $(z/\ell)^2\D \hat s^2_{\rm LeBrun}$ is self-dual and Einstein. In the following analysis, we set $\ell =1$ 
and try to identify $\D \hat s^2_{\rm LeBrun}$ with $\D \hat s^2_{-}=\Phi \D s_c^2$.  

\subsection{Hyperbolic type}

Let us get started with the $y$-independent LeBrun metric (\ref{LeBrun}). 
Performing the coordinate transformation
\begin{align}
\label{tyzx_hyperbolic}
&t=\tau \,, \qquad y=\sigma+b \tau \,,\qquad 
z=\frac{1+bpq}{p+q}\,, \notag \\
& \D x=\frac{1-bq^2}{Q(q)}\D q -\frac{1-bp^2}{P(p)}\D p\,,
\end{align}
where $b$ is a constant and 
\begin{align}
\label{Huo_PD}
H&=\frac{(p+q)^3(q-p)}{(1-bq^2)^2P(p)+(1-bp^2)^2Q(q)}\,,  \quad 
e^u=\frac{P(p)Q(q)}{(p+q)^4} \,, \notag \\
\omega&=\frac{q^2(1-bq^2) P(p)+p^2(1-bp^2) Q(q)}{(1-bq^2)^2P(p)+(1-bp^2)^2Q(q)}\D y \,,
\end{align}
one obtains the ambi-toric K\"ahler metric $\D \hat s^2_-=\Phi \D s_c^2$ of hyperbolic type (\ref{hyperbolic}). 
This coordinate transformation is insensitive to 
the precise form of structure functions $P(p)$ and $Q(q)$. 
Now let us require that the conformally transformed metric $\D s^2=z^2\D \hat s_-^2$ is 
Einstein and its Weyl curvature is self-dual. This restricts $z $ to be 
(\ref{tyzx_hyperbolic}) and  the structure functions to take the form 
\begin{align}
\label{PQ_PD}
P(p)&=a_0+a_1 p+a_2 p^2+ba_1 p^3 +(b^2a_0-1)p^4\,, \notag \\
Q(q)&=-a_0+a_1 q-a_2 q^2+b a_1q^3+(1-b^2a_0)q^4 \,,
\end{align}
where $a_{0,1,2}$ are constants. 
It follows that  the conformal transformation (\ref{PT}) gives rise to the $y$-independent Przanowski-Tod metric. An application of  the result in section~\ref{sec:sd} immediately concludes that the solution  admits two independent Killing spinors. 

The obtained self-dual Einstein metric is nothing but the Euclidean Pleba\'nski-Demia\'nski solution~\cite{Apostolov:2013oza,Nozawa:2015qea}. An interesting aspect of this solution is that it has a description in terms of the Carter family~\cite{Carter:1968ks}, which is a $b=0$ limit of the Pleba\'nski-Demia\'nski solution. Consequently, an acceleration parameter $b$ can be gauged away by a suitable coordinate transformation $(\tau, q,p,\sigma)\mapsto (\ti\tau, \ti q,\ti p,\ti\sigma)$~\cite{Nozawa:2015qea}. To show this, the Killing-Yano tensor played a central  role. However, the Killing-Yano tensor is inessential, as far as the existence of the second Killing spinor is concerned. 
The Killing spinors constructed here are exhaustive, since the upper bound on the number of Killing vectors in the self-dual Carter metric is two~\cite{Houri:2014hma}, which are the ones built out of the Killing spinors $\epsilon, \ti \epsilon$. 

The solution to the Toda system (\ref{Huo_PD}) is given implicitly as functions of ($x, z$). The relation will be more manifest if one can find an algebraic equation of $u$ in terms of ($x,z$), as discussed in \cite{Plansangkate:2016mod} for the ungauged theory. The basic idea behind this is to write the solution to the Toda system in terms of an axially symmetric harmonic function on $\mathbb E^3$.  
Contrary to \cite{Plansangkate:2016mod} where the harmonic function is given in advance, one needs to solve the inverse relations (c.f, eq. (81)--(84) in \cite{Nozawa:2015qea}) to obtain the harmonic function in the present case. This is not straightforward but an interesting direction for further understanding of self-dual solutions. 

\subsection{Parabolic type}

Consider the LeBrun space and the following coordinate transformation
\begin{align}
\label{zxty_parabolic}
z=p+q\,, \quad \D x=\frac{\D p}{P(p)}-\frac{\D q}{Q(q)} \,,\quad 
t=\tau \,, \quad y=\sigma \,,
\end{align}
with 
\begin{align}
\label{Hou_parabolic}
H&=\frac{q-p}{P(p)+Q(q)}  \,, \qquad 
e^u=P(p)Q(q) \,, \notag \\
\omega&=\frac{qP(p)+p Q(q)}{P(p)+Q(q)} \D y\,.
\end{align}
One thus gets the ambi-K\"ahler metric $\D \hat s^2_-=\Phi \D s_c^2$ of parabolic type.
The parabolic type ambi-K\"ahler metric $\D \hat s^2_-$ is also referred to as the orthotoric metric
and attracts some attention from the geometric point of view, 
because it admits the Hamiltonian two-form~\cite{Apostolov:2006gra} and the Killing-Yano tensor with a three-form torsion~\cite{Houri:2012eq}. 

Imposing the Toda system to (\ref{zxty_parabolic}), (\ref{Hou_parabolic}), we find 
\begin{align}
\label{}
P(p)&=a_0+a_1p+a_2 p^2 \,, \notag \\
Q(q)&=-a_0+(2+a_1)q-a_2 q^2 \,,
\end{align}
where $a_{0,1,2}$ are constants, and the allowed conformal factor $z$ is only (\ref{zxty_parabolic}). 
The conformal transformation (\ref{PT}) then gives rise to the $y$-independent 
Przanowski-Tod space and two Killing spinors exist.

Since the expression of ($\ti x, \ti z$) in terms of ($q,p $) is fairly arkward, 
it is desirable to find a more convenient coordinate system to make further progress, by choosing 
$c_1$ and $c_2$ appropriately. 
Here we observe that the following two-form is the Killing-Yano tensor 
\begin{align}
\label{KYortho}
f=f_+ e^1\we e^2+f_0 (e^1 \we e^3 +e^2\we e^4) +f_- e^3 \we e^4 \,, 
\end{align}
satisfying $\nabla_{(\mu} f_{\nu)\rho} =0$, where 
\begin{align}
\label{}
f_0=&-\frac{2\sqrt{P(p)Q(q)}}{p+q}\,, 
\notag \\
f_\pm =&\frac{q-p\pm [2a_0-a_1q+(2+a_1)p-2a_2 pq]}{q+p} \,. 
\end{align}
Here we worked in the orthonormal frame 
\begin{align}
e^1=&\sqrt{\frac{Q(q)}{q-p}}\frac{\D \tau+p \D \sigma}{q+p}\,, ~~
e^2=\sqrt{\frac{q-p}{Q(q)}}\frac{\D q}{q+p} \,,\notag \\
e^3=&\sqrt{\frac{q-p}{P(p)}}\frac{\D p}{q+p} \,, ~~
e^4=\sqrt{\frac{P(p)}{q-p}}\frac{\D \tau+q\D \sigma}{q+p}\,. 
\label{conformalortho}
\end{align}
According to the analysis in \cite{Houri:2007xz,Krtous:2008tb}, the four dimensional spaces admitting the Killing-Yano tensor are exhausted by the Carter family~\cite{Carter:1968ks} corresponding to the $b= 0$ limit of the Pleba\'nski-Demia\'nski solution. This means that a suitable change of basis brings the metric (\ref{conformalortho}) into the Carter form. A key role to this aim is played by the the Killing-Yano tensor (\ref{KYortho}). Since two eigenvalues of $f$ are nondegenerate, we are able to employ them  as new coordinates 
\begin{align}
\label{}
\ti q&=\frac 12 \left(\sqrt{(f_+-f_-)^2+4f_0^2}+f_++f_-\right)\,, \notag \\
\ti p&=\frac 12 \left(\sqrt{(f_+-f_-)^2+4f_0^2}-(f_++f_-)\right)\,.
\end{align}
Preliminary computations show that the metric (\ref{conformalortho}) can be cast into the form
\begin{align}
\label{}
\D s^2=&\frac{\ti Q(\ti q)}{\ti q^2-\ti p^2}(\D \ti \tau+\ti p^2\D \ti \sigma)^2
+(\ti q^2-\ti p^2)\left(\frac{\D \ti q^2}{\ti Q(\ti q)}
+\frac{\D \ti p^2}{\ti P(\ti p)}\right)\notag \\
&+\frac{\ti P(\ti p)}{\ti q^2-\ti p^2}(\D \ti \tau+\ti q^2\D \ti \sigma)^2\,,
\end{align}
where
\begin{align}
\label{}
\ti \tau =\frac{(1+a_1)^2-4a_0a_2}{16a_0} \tau -\frac 14 \sigma 
\,, \quad \ti \sigma=-\frac {\tau}{16a_0} \,,
\end{align}
and the structure functions are given by 
\begin{align}
\label{}
\ti P(p)&=\ti a_0+\ti a_1\ti p+\ti a_2\ti p^2-\ti p^4 \,, \notag   \\
\ti Q(q)&=-\ti a_0+\ti a_1\ti q-\ti a_2\ti q^2+\ti q^4 \,,
\end{align}
with 
\begin{align}
\label{}
\ti a_0&=  (1-a_1^2+4a_0a_2)[(a_1+1)(a_1+3)-4a_0a_2]\,, \notag \\
\ti a_1&=8(1+a_1) \,, \quad \ti a_2=2(a_1^2+2a_1+3-4a_0a_2)\,.
\end{align}
This is the self-dual Carter metric ($b=0$ limit of (\ref{PQ_PD})), as we desired to show. 
The coordinate transformation to the Carter metric corresponds to the choice 
$c_1=0, c_2=-4$ in (\ref{dxt}).

\subsection{Elliptic type}

Our last example is the elliptic type, for which the desired transformation from the 
LeBrun space is given by 
\begin{align}
\label{zxty_elliptic}
&z=\frac{1-pq+b(p+q)}{1+pq}\,,\quad t=\tau \,,\quad y=\sigma-b \tau \,, \notag \\
&\D x=-\frac{2p+b(p^2-1)}{2P(p)}\D p+
\frac{2q+b(q^2-1)}{2Q(q)} \D q\,, 
\end{align}
where $b$ is a constant with 
\begin{align}
\label{}
H&=\frac{(q-p)(1+p q)^3}{[2q+b(q^2-1)]^2P(p)+[2p+b(p^2-1)]^2Q(q)} \,,\notag \\
e^u&=\frac{4P(p)Q(q)}{(1+p q)^4} \,, \\
\omega &=\frac{\varpi \D y}
{[2q+b(q^2-1)]^2P(p)+[2p+b(p^2-1)]^2Q(q)} \,. \notag 
\end{align}
Here $\varpi=(q^2-1)[2q+b(q^2-1)]P(p)+(p^2-1)[2p+b(p^2-1)] Q(q)$. 
The Toda system is only compatible with (\ref{zxty_elliptic}) and  quartic functions of the form
\begin{align}
\label{}
P(p)=\sum_{j=0}^4 a_j p^j\,,\quad 
Q(q)=\sum_{j=0}^4 (-1)^{j+1}a_{4-j} q^j \,, 
\end{align}
where the constants $a_j$ are subjected to the constraints 
\begin{align}
\label{}
a_1+a_3+2b(a_4-a_0)&=0\,, \notag \\
a_1(1-b^2)+b(a_2-2a_0)+b^3(a_0-a_4)&=1\,,
\end{align}
It follows that the elliptic type metric falls into the $y$-independent Przanowski-Tod space. 

Also in this case, it turns out that the metric can be reduced to the self-dual Carter family. 
Combined with the results in  \cite{Houri:2007xz,Krtous:2008tb}, this can be immediately appreciated from the existence of the Killing-Yano tensor. 
Now, adopting the frame 
\begin{align}
\label{}
e^1=&\sqrt{\frac{Q(q)}{(q-p)(1+pq)}}\frac{2p\D \tau+(p^2-1)\D \sigma}{1-pq+b(p+q)}\,, \notag\\
e^2=&\sqrt{\frac{(q-p)(1+pq)}{Q(q)}}\frac{\D q}{1-pq+b(p+q)}\,, \notag \\
e^3=&\sqrt{\frac{(q-p)(1+pq)}{P(p)}}\frac{\D p}{1-pq+b(p+q)}\,, \notag \\
e^4=&\sqrt{\frac{P(p)}{(q-p)(1+pq)}}\frac{2q\D \tau+(q^2-1)\D \sigma}{1-pq+b(p+q)}\,, 
\end{align}
the Killing-Yano tensor takes exactly the same form as (\ref{KYortho}), 
where $f_0, f_\pm $ are replaced in the present case by
\begin{align}
\label{}
f_0&=-\frac{2(1+b^2)\sqrt{P(p)Q(q)}}{(1+pq)[1-pq+b(p+q)]}\,, \notag \\
f_\pm &=\frac{q-p \pm [b(1+pq)]^{-1}\ti f }{1-pq+b(p+q)}\,,
\end{align}
with 
\begin{align}
\label{}
\ti f=&1-2b^2pq +p^2q^2+(1+b^2)[(1+bp)(q-b)qa_1
\notag \\&
+\{1+q^2(1+p^2)+bp(1-pq)+b^2pq\}a_3\notag \\
&+2b(1+p^2)(1+q^2)a_4
] \,.
\end{align}
Choosing the eigenvalues of the Killing-Yano tensor as new coordinates, 
one finds that the self-dual elliptic ambi-toric metric can be transformed in the 
Carter form, although we do not attempt to do this here. 

To summarize, when three classes of regular ambi-toric K\"ahler manifold
are taken as the LeBrun space, the corresponding Przanowski-Tod spaces are all 
degenerate into the Carter form. Since all the three cases belong to Petrov-type D~\cite{Apostolov:2013oza}, 
only a single eigenvalue of the Weyl tensor is independent for the self-dual solution, which is a  principal ground for degeneracy.  A somewhat more surprising issue is that one can write the metric into the Carter form. This is ascribed to the existence of the closed conformal Killing-Yano tensor~\cite{Houri:2007xz,Krtous:2008tb} corresponding to the Hodge dual to the Killing-Yano tensor. For the Przanowski-Tod metric of the form 
$\D s^2=z^{-2}\D \hat s_-^2=(\Phi/z)^2 \D \hat s_+^2$, this closed conformal Killing-Yano tensor can be constructed out of two kinds of the degenerate conformal Killing-Yano tensors, i.e, 
$(\Phi/z)\hat \Omega_+$ and the twistor two-form (\ref{twistor}). Since the former conformal Killing-Yano tensor is present exclusively in the ambi-K\"ahler geometry, a description in Carter form is unlikely to occur for the general $y$-independent Przanowski-Tod space. 

\bigskip

\section{Summary and closing remarks}
\label{remark}

In this paper, we have studied the Euclidean supersymmetry of the 
self-dual solutions in the framework of  ${\cal N}=2$ minimal gauged supergravity in four dimensions. 
If the Przanowski-Tod space admits the second Killing vector of the form $\partial/\partial y$, 
there exists a one-parameter family of writing the metric into the Przanowski-Tod form. 
This demonstrates that there exists another Killing spinor, whose instanton field strength is
distinct from the original one.  Our analysis was geared to providing useful tools to check the supersymmetry for the wide variety of self-dual solutions with two commuting isometries, and will be 
profitable when we try to explore the gauge theory in curved background.

We presumed throughout the paper that the gauge field $A_\mu$ and the coupling constant $\ell^{-1}$
are both real. In Euclidean signature, these restrictions are not necessary in general~\cite{Dunajski:2010zp,Klemm:2015mga}. It would be nice if a similar mechanism works even if
we relax these conditions.

Using the scheme developed in section \ref{sec:sd}, we discussed the supersymmetry of the self-dual Reissner-Nordstr\"om-Taub-NUT-AdS metric. To the best of our knowledge, the Killing spinor discovered here is new, although this subject has been studied for many decades~\cite{Martelli:2012sz,Farquet:2014kma,AlonsoAlberca:2000cs,Page:1985bq}. We answered in the affirmative fashion that the Reissner-Nordstr\"om-Taub-NUT-AdS indeed has two linearly independent Killing spinors with a one-parameter family of field strength when the Weyl tensor is self-dual.

We found a curious property that three types of self-dual Einstein metrics built out of the regular ambi-toric K\"ahler class are all transmuted into the self-dual Carter family, owing to the Killing-Yano tensor. This interesting feature does not show up in the Lorentzian signature, because the type-D counterparts of parabolic and elliptic types do not exist and there is no concept of self-duality. 
The ambi-toric geometry yields the independent interest about the 
base space for the supersymmetric solutions to five dimensional gauged supergravity~\cite{Cassani:2015upa,Chimento:2016mmd}. The elliptic type ambi-K\"ahler metric has not been analyzed so extensively thus far, in contrast to the hyperbolic and parabolic cases. It seems interesting to elucidate if the non-self-dual elliptic class admits some linear tensorial fields analogous to the Hamiltonian 2-form~\cite{Apostolov:2006gra} and the Killing-Yano tensor possibly with a torsion~\cite{Houri:2012eq,Chervonyi:2015ima}. We intend to pursue these issues in the near future.

\acknowledgments
The author is indebted to Takahiro Tanaka for discussions.
This work was partially supported by MEXT Grant-in-Aid for Scientific Research on Innovative Areas ``New Developments in Astrophysics Through Multi-Messenger Observations of Gravitational Wave Sources" (Grant Number A05 24103006).


\end{document}